\documentclass[11pt,a4paper,english,superscriptaddress,aps,preprint]{revtex4}
\usepackage{xcolor}
\usepackage{amsfonts}
\usepackage{graphics}
\usepackage{graphicx}
\usepackage{hyperref}
\usepackage{slashed}
\usepackage{amsmath}
\usepackage{cleveref}
\usepackage{amssymb}
\makeatletter
\usepackage{babel}
\newcommand{\bea}{\begin{eqnarray}}
\newcommand{\eea}{\end{eqnarray}}

\begin{document}

\title{Dualization between supersymmetric self-dual and topologically massive models coupled to matter in four-dimensional superspace}

\author{F. S. Gama}
\email{fisicofabricio@yahoo.com.br}
\affiliation{Departamento de F\'{\i}sica, Universidade Federal da Para\'{\i}ba\\
 Caixa Postal 5008, 58051-970, Jo\~ao Pessoa, Para\'{\i}ba, Brazil}

\author{R. V. Maluf}
\email{r.v.maluf@fisica.ufc.br}
\affiliation{Universidade Federal do Cear\'a (UFC), Departamento de F\'isica,\\ Campus do Pici, Fortaleza - CE, C.P. 6030, 60455-760 - Brazil.}

\author{J. R. Nascimento}
\email{jroberto@fisica.ufpb.br}
\affiliation{Departamento de F\'{\i}sica, Universidade Federal da Para\'{\i}ba\\
 Caixa Postal 5008, 58051-970, Jo\~ao Pessoa, Para\'{\i}ba, Brazil}

\author{A. Yu. Petrov}
\email{petrov@fisica.ufpb.br}
\affiliation{Departamento de F\'{\i}sica, Universidade Federal da Para\'{\i}ba\\
Caixa Postal 5008, 58051-970, Jo\~ao Pessoa, Para\'{\i}ba, Brazil}

\author{P. Porfirio}
\email{pporfirio@fisica.ufpb.br}
\affiliation{Departamento de F\'{\i}sica, Universidade Federal da Para\'{\i}ba\\
	Caixa Postal 5008, 58051-970, Jo\~ao Pessoa, Para\'{\i}ba, Brazil}

\begin{abstract}
Using the superfield formalism and the master action approach, we prove, both at the classical and quantum levels, the dual equivalence between four-dimensional supersymmetric self-dual and topologically massive models coupled to dynamical matter.
 \end{abstract}

\maketitle

\section{Introduction}

The concept of symmetry is a powerful tool for interpreting physical phenomena. A fundamental class of symmetries is  presented by the duality symmetries. The primary notion of duality consists of connecting different theories or opposing regimes of the same model, each containing different associated symmetries. In fact, whether different theories are dual it suggests that they represent themselves as a manifestation of the same theory with different ``guises". In addition, dual properties provide a powerful mechanism to seek out and understand non-perturbative effects within the context of quantum field theories \cite{Alvarez-Gaume:1997dpg} and condensed matter systems \cite{Zaanen:2015oix}.

Thus, the study of dualities is an ever-present and relevant topic in physics.  Dualities play an important role in the context of toroidal compactifications \cite{Green:1982sw} and string/M-theory. Namely, the five string theories we have known so far are -- albeit at first sight they look radically different --
 linked together by dualities. For example, the duality between type IIA and type IIB string theories is called T-duality
which is related to the geometrical properties of the target space(see \cite{Alvarez:1994dn, tduality} for a review), while type I and $SO(32)$ heterotic string
theories are related to each other by S-duality which means that one theory in the strong coupling regime is equivalent 
to the other in the weak coupling regime (see \cite{Polchinski} for a review on this subject). Another kind of special duality is the so-called gauge/gravity duality which in turn relates two theories with different natures -- a (super)gravity theory to a gauge theory. A particular realization of this duality is the $AdS/CFT$ correspondence that links a low-energy string theory in $AdS$ spacetime to a strong coupling regime of a conformal field theory on {$AdS$} boundary \cite{maldacena}. Besides of this, among duality processes, bosonization possesses a special importance being widely used to investigate non-perturbative properties in low-dimensional condensed matter systems \citep{BurgessQuevedo}. In $1+1$ dimensions, it is possible to establish a fermion-boson correspondence based on the properties of Fermi surfaces \cite{Mandelstam}. This duality can be generalized for non-abelian fields \cite{Witten84} and higher dimensions \cite{Marino, Burgess}. Recently, bosonization has led to new $2+1$ connections called web duality \cite{Seiberg:2016gmd,Hernaski1,Hernaski2}.

A well-known example of duality involves topologically massive gauge theories \cite{Dunne:1995ai}. The first discovered case relates the self-dual (SD) \cite{auto dual original} and Maxwell-Chern-Simons (MCS) \cite{MCS auto dual} models. These two theories describe a single massive particle of spin-1 in $2+1$ flat spacetime. However, only the MCS model is gauge-invariant. The equivalence between the SD and MCS models was initially established by Deser and Jackiw \cite{MCS auto dual}, and since then several studies of this equivalence were carried out in the literature \cite{Fradkin,Bralic,Banerjee,Banerjee2,malacarne,Minces,Anacleto2001}. Particularly, considering couplings with fermionic fields, it was shown in \cite{malacarne} that the models are equivalent, as long as a Thirring interaction is included. Furthermore, supersymmetric \cite{Karlhede2,Ferrari1,Ferrari2} and non-commutative \cite{Gomes,HarikumarRivelles} extensions to the duality involving the SD and MCS models were studied in several contexts.

The Chern-Simons term plays a key role at the heart of the SD-MCS duality. An alternative topological term in $3+1$ dimensions can be formed from a gauge vector field $A_{\mu}$ and a rank-2 antisymmetric tensor field $B_{\mu\nu}$, also called the Kalb-Ramond field \cite{KalbRamon}. The associated topological term is known as the $B \wedge F$ term \cite{BFOriginal,Banerjee3,MKR,Oda,Harikumar,Lahiri}. Hence, a higher-dimensional  generalization of the MCS model consists of the Maxwell and Kalb-Ramond fields coupled by a $B \wedge F$ term. The Maxwell-Kalb-Ramond (MKR) theory is gauge invariant, unitary, and renormalizable when minimally coupled to fermions and represents a massive spin-1 particle \cite{BFOriginal}. The Kalb-Ramond field appears in several contexts in the literature, in particular we can mention its study within contexts of string theories \cite{Soo}, quantum field theory \cite{Deguchi1999, Hari}, supersymmetry \cite{Almeida},  Lorentz symmetry violation \cite{Altschul2010,Hernaski2016,Maluf2018,Mariz2019}, black hole solutions \cite{Euclides2020}, cosmology \cite{Grezia,Maluf:2021eyu}, and brane words scenarios \cite{Wilami1,Wilami2}.

A self-dual version of the MKR model ($SD_{BF}$) was first studied in Ref. \cite{BF_dual}, where the authors included the $B \wedge F$ term added of gauge non-invariant mass terms. In this work, the gauge embedding procedure \cite{Anacleto2001} was used to prove the classical equivalence between the models without and in the presence of fermion couplings. Analogous to the SD/MCS case in $2+1$ dimensions, interactions with fermionic fields require Thirring-like terms to preserve the duality mapping. More recently, the proof of duality has been extended to the quantum level through the master action approach and considering arbitrary non-conserved matter currents \cite{Maluf:2020fch}. 

The main objective of the present work is to obtain a supersymmetric generalization of the duality between $SD_{BF}$ and MKR theories, both at classical and quantum levels, employing their superfield description. Throughout this paper, we use the definitions and conventions for the superfield formalism adopted in \cite{GGRS} (see also \cite{Siegel} where the chiral spinor superfield has been introduced originally) and further applied in \cite{Almeida}, where the supersymmetric theory of an antisymmetric tensor field, described within the superfield approach by a chiral spinor superfield, has been studied in the one-loop approximation (see also \cite{firstBF} where a simplified version of this theory was considered). Actually, this paper can be treated as a natural continuation of \cite{Almeida}.

This work is organized as follows. In the section 2, we discuss the equivalence of self-dual and topologically massive theories at the tree level, both for the case of free models and of the presence of nontrivial matter couplings. In the section 3, we demonstrate the equivalence of these theories at the perturbative level, through performing integration over dynamical fields. Finally, in the Summary we discuss our results.

\section{Classical Equivalence}

In this section, for the sake of convenience, we firstly investigate the classical equivalence in the case of free dual theories. Then, we investigate the classical equivalence in the case of the dual theories interacting with chiral matter superfields. Earlier their equivalence has been discussed in \cite{BuchKuz}, where, however, a completely distinct methodology has been used while we follow the master action approach developed in \cite{MCS auto dual} known as a standard tool for studying the duality (for its further development see f.e. \cite{Anacleto2001}). Moreover, in \cite{BuchKuz} the real and spinor scalar superfields were couplied to a curved background while we consider the flat superspace only, but, unlike \cite{BuchKuz}, our superfields are coupled to a chiral matter.

\subsection{Free theories}

In $\mathcal{N}=1$, $d=4$ superspace, the supersymmetric self-dual model is formulated in terms of a real superfied $Z$ and a chiral spinor superfield $\pi^\alpha$, which are not subject to gauge transformations. This model is described by the following first-order action \cite{Siegel}:
\bea
\label{sd_action}
S_{SD}=-\frac{\chi\vartheta}{2}\int d^8z\left(\pi^\alpha D_\alpha Z+\bar{\pi}^{\dot{\alpha}} \bar{D}_{\dot{\alpha}} Z\right)+\frac{1}{4}\left(\int d^6z\pi^\alpha \pi_\alpha+\int d^6\bar z\bar\pi^{\dot{\alpha}}\bar\pi_{\dot{\alpha}}\right)+\frac{m^2}{2}\int d^8zZ^2,
\eea
where $m$ is a constant with mass dimension equal to $1$, $\vartheta$ is a dimensionless constant, and $\chi=\pm 1$ determines
if the model is self-dual or anti-self-dual. 
The theory (\ref{sd_action}) is a supersymmetric extension of the self-dual $B\wedge F$ model \cite{BF_dual} and describes massive superfields which can be seen as follows. The Euler-Lagrange equations following from (\ref{sd_action}) have the forms:
\bea
\label{eq_Z}
\frac{\delta S_{SD}}{\delta Z}&=&\frac{\chi\vartheta}{2}\left(D^\alpha\pi_\alpha+\bar{D}^{\dot{\alpha}}\bar{\pi}_{\dot{\alpha}}\right)+m^2Z=0;\\
\label{eq_pi}
\frac{\delta S_{SD}}{\delta \pi^\alpha}&=&-\frac{\chi\vartheta}{2}\bar{D}^2D_\alpha Z+\frac{1}{2}\pi_\alpha=0\\
\label{eq_barpi}
\frac{\delta S_{SD}}{\delta\bar\pi^{\dot{\alpha}}}&=&-\frac{\chi\vartheta}{2}D^2\bar{D}_{\dot{\alpha}} Z+\frac{1}{2}\bar{\pi}_{\dot{\alpha}}=0.
\eea
We can use well-known properties of the covariant derivatives $D_\alpha$ and $\bar{D}_{\dot{\alpha}}$ along with the above equations of motion to show that
\bea
\label{const}
D^2 Z = 0 \ ; \ \bar{D}^2 Z = 0 \ ; \ D^\alpha \pi_\alpha = \bar{D}^{\dot{\alpha}}\bar{\pi}_{\dot{\alpha}}.
\eea
With the help of these constraints, we can manipulate Eqs. (\ref{eq_Z}-\ref{eq_barpi}) to decouple the superfields and derive
\bea
\left(\Box-\frac{m^2}{\vartheta^2}\right)Z=0 \ ; \ \left(\Box-\frac{m^2}{\vartheta^2}\right)\pi^\alpha=0 \ ; \ \left(\Box-\frac{m^2}{\vartheta^2}\right)\bar{\pi}^{\dot{\alpha}}=0.
\eea
Therefore, we can conclude that $Z$ and $\pi^\alpha$ satisfy Klein-Gordon equations with the mass square $\frac{m^2}{\vartheta^2}$. Indeed, the algebraic Euler-Lagrange equations (\ref{eq_pi}) and (\ref{eq_barpi}) can be solved for $\pi_\alpha$ and $\bar{\pi}_{\dot{\alpha}}$, and the substitution of these solutions into (\ref{sd_action}) gives the massive vector multiplet model. On the other hand, solving (\ref{eq_Z}) for $Z$, this superfield can be eliminated from the action (\ref{sd_action}) altogether yielding the massive tensor multiplet model \cite{Siegel}.

Contrastingly, the supersymmetric extension of the topologically massive model \cite{MKR} is formulated in terms of a real prepotential $V$ and a chiral spinor prepotential $\psi^\alpha$, which are subject to gauge transformations
\bea
\label{gaugetrans}
\delta V=i(\bar\Lambda-\Lambda) \ ; \ \delta\psi_\alpha=i\bar D^2D_\alpha L \ ; \ \delta\bar\psi_{\dot{\alpha}}=-iD^2\bar D_{\dot{\alpha}}L,
\eea
and correspond to the following gauge-invariant superfield strengths
\bea
\label{WG}
W_\alpha=i\bar D^2D_\alpha V \ ; \ \bar{W}_{\dot{\alpha}}=-iD^2\bar{D}_{\dot{\alpha}} V \ ; \  G=-\frac{1}{2}(D^\alpha\psi_\alpha+\bar D^{\dot{\alpha}}\bar\psi_{\dot{\alpha}}).
\eea
The second-order model which describes a massive gauge
theory consists of quadratic terms in the superfield strengths and a term which the variation yields a total divergence \cite{Siegel}:
\bea
\label{tm_action}
S_{TM}=\frac{\vartheta^4}{4}\left(\int d^6zW^\alpha W_\alpha+\int d^6\bar{z}\bar{W}^{\dot{\alpha}}\bar{W}_{\dot{\alpha}}\right)-\frac{\vartheta^2}{2m^2}\int d^8zG^2-\chi\vartheta\int d^8zVG.
\eea
In order to establish the duality between the self-dual and topologically massive models, let determine the Euler-Lagrange equations from the action (\ref{tm_action}). They have the form
\bea
\label{eq_V}
\frac{\delta S_{TM}}{\delta V}&=&-\frac{i\vartheta^2}{2}D^\alpha W_\alpha+\frac{i\vartheta^2}{2}\bar{D}^{\dot{\alpha}} \bar{W}_{\dot{\alpha}}-\chi\vartheta G=0;\\
\label{eq_psi}
\frac{\delta S_{TM}}{\delta \psi^\alpha}&=&-\frac{\vartheta^2}{2m^2}\bar D^2D_\alpha G+i\frac{\chi\vartheta}{2}W_\alpha=0;\\
\label{eq_barpsi}
\frac{\delta S_{TM}}{\delta\bar\psi^{\dot{\alpha}}}&=&-\frac{\vartheta^2}{2m^2}D^2\bar D_{\dot{\alpha}} G-i\frac{\chi\vartheta}{2}\bar W_{\dot{\alpha}}=0.
\eea
It is possible to manipulate these equations, using the properties of the spinor covariant derivatives and the superfield strengths, to show that the superfield strengths also satisfy Klein-Gordon equations with the squared mass $\frac{m^2}{\vartheta^2}$:
\bea
\label{kg_Ws}
\left(\Box-\frac{m^2}{\vartheta^2}\right)W^\alpha=0 \ ; \ \left(\Box-\frac{m^2}{\vartheta^2}\right)\bar{W}^{\dot{\alpha}}=0 \ ; \ \left(\Box-\frac{m^2}{\vartheta^2}\right)G=0.
\eea
A direct inspection of the two different sets of equations of motion (\ref{eq_Z}-\ref{eq_barpi}) and (\ref{eq_V}-\ref{eq_barpsi}) unveils that the equations of motion for the superfields $\{Z,\pi^\alpha,\bar{\pi}^{\dot{\alpha}}\}$ looks like those for the superfield strengths $\{W^\alpha,\bar{W}^{\dot{\alpha}},G\}$. This shows that (\ref{eq_Z}-\ref{eq_barpi}) are identical with (\ref{eq_V}-\ref{eq_barpsi}) when the identifications
\bea
Z\longleftrightarrow-\frac{\chi\vartheta}{m^2}G \ ; \ \pi_\alpha\longleftrightarrow-i\chi\vartheta W_\alpha \ ; \ \bar{\pi}_\alpha\longleftrightarrow i\chi\vartheta\bar{W}_{\dot{\alpha}}
\eea
are used. Therefore, there is a classical equivalence between the supersymmetric self-dual and topologically massive models, that is, equivalence at the level of equations of motion.

Alternatively, we can also establish the duality discussed above by means of the master action approach \cite{MCS auto dual}. The master action approach is useful because it reveals the common origin of dual theories such as (\ref{sd_action}) and (\ref{tm_action}). Besides, the master action is a valuable theoretical tool for establishing the equivalence of generating functionals of dual theories at the quantum level.

To obtain the master action, we introduce a set of gauge-invariant auxiliary superfields $\{Z,\pi^\alpha,\bar{\pi}^{\dot{\alpha}}\}$ in order to reduce the number of derivatives in the original
second-order model (\ref{tm_action}). Doing this, we obtain the following gauge-invariant master action
\bea
\label{master}
S_M &=& \frac{1}{4}\left(\int d^6z\pi^\alpha \pi_\alpha+\int d^6\bar z\bar\pi^{\dot{\alpha}}\bar\pi_{\dot{\alpha}}\right)+i\frac{\chi\vartheta}{2}\left(\int d^6z\pi^\alpha W_\alpha-\int d^6\bar z\bar\pi^{\dot{\alpha}}\bar W_{\dot{\alpha}}\right)\nonumber\\
&+&\int d^8z\left(\frac{m^2}{2} Z^2+\chi\vartheta ZG\right)-\chi\vartheta\int d^8zVG.
\eea
Let us demonstrate that (\ref{master}) is indeed a first-order formulation of the supersymmetric topologically massive model. Varying $S_M$ with respect to the auxiliary superfields, we find
\bea
\label{master_eqZ}
\frac{\delta S_{M}}{\delta Z}&=&m^2Z+\chi\vartheta G=0;\\
\label{master_eq_pi}
\frac{\delta S_{M}}{\delta \pi^\alpha}&=&\frac{1}{2}\pi_\alpha+i\frac{\chi\vartheta}{2}W_\alpha=0;\\
\label{master_eq_barpi}
\frac{\delta S_{M}}{\delta\bar\pi^{\dot{\alpha}}}&=&\frac{1}{2}\bar{\pi}_{\dot{\alpha}}-i\frac{\chi\vartheta}{2}\bar{W}_{\dot{\alpha}}=0.
\eea
Solving these algebraic equations for the auxiliary superfields and substituting them into (\ref{master}), we find that $S_M[V,\psi^\alpha,\bar{\psi}^{\dot{\alpha}}]=S_{TM}$, thus demonstrating the equivalence of (\ref{master}) with (\ref{tm_action}).

On the other hand, in order to prove that (\ref{master}) is also equivalent to (\ref{sd_action}), let us vary $S_M$ with respect to the gauge prepotentials and get the following equations of motion
\bea
\label{master_eq_V}
\frac{\delta S_{M}}{\delta V}&=&\frac{\chi\vartheta}{2}\left(D^\alpha\pi_\alpha+\bar{D}^{\dot{\alpha}}\bar{\pi}_{\dot{\alpha}}\right)-\chi\vartheta G=0;\\
\label{master_eq_psi}
\frac{\delta S_{M}}{\delta \psi^\alpha}&=&\frac{\chi\vartheta}{2}\bar{D}^2D_\alpha Z-\frac{\chi\vartheta}{2}\bar{D}^2D_\alpha V=0;\\
\label{master_eq_barpsi}
\frac{\delta S_{M}}{\delta\bar\psi^{\dot{\alpha}}}&=&\frac{\chi\vartheta}{2}D^2\bar{D}_{\dot{\alpha}} Z-\frac{\chi\vartheta}{2}D^2\bar{D}_{\dot{\alpha}} V=0.
\eea
These equations can be immediately integrated and give the following general solutions
\bea
\label{solutions}
V = Z+i\left(\bar{\Psi}-\Psi\right) \ ; \ \psi_\alpha=-\pi_\alpha+i\bar{D}^2D_\alpha N \ ; \ \bar{\psi}_{\dot{\alpha}}=-\bar{\pi}_{\dot{\alpha}}-iD^2\bar{D}_\alpha N,
\eea
where $\Psi$ and $N$ are arbitrary chiral and real scalar superfields, respectively.
  
Finally, plugging these solutions into (\ref{master}), we completely eliminate the gauge prepotentials from the master action and show that $S_M[Z,\pi^\alpha,\bar{\pi}^{\dot{\alpha}}] = S_{SD}$, which implies that (\ref{master}) is also equivalent to (\ref{sd_action}).

Accordingly, we can conclude that the models (\ref{sd_action}) and (\ref{tm_action}) originate from a single action (\ref{master}). This demonstrates the classical equivalence between the supersymmetric self-dual and topologically massive models, that is, these models are different mathematical descriptions of the same classical physics.

\subsection{Interaction with chiral matter superfields}

In the previous subsection, we investigated the classical equivalence in the case of free theories. However, fields in the real world interact with each other. Thus, we now turn to the study of the classical equivalence of interacting theories, which are more realistic and interesting descriptions of the physical phenomena. We will do this including new terms in the master action that will couple different superfields to each other. In particular, for the sake of simplicity, we will include in the master action (\ref{master}) only linear couplings with external superfields, whose associated currents are composed only of dynamical chiral matter superfields, which we generically denote by $\Phi$. Thus, we define the new master action as
\bea
\label{master_1}
S_M^{(1)} &=& \frac{1}{4}\left(\int d^6z\pi^\alpha \pi_\alpha+\int d^6\bar z\bar\pi^{\dot{\alpha}}\bar\pi_{\dot{\alpha}}\right)+i\frac{\chi\vartheta}{2}\left(\int d^6z\pi^\alpha W_\alpha-\int d^6\bar z\bar\pi^{\dot{\alpha}}\bar W_{\dot{\alpha}}\right)\nonumber\\
&+&\int d^8z\left[\frac{m^2}{2} Z^2+\chi\vartheta (Z-V)G\right]+\int d^6z\pi^\alpha j_\alpha+\int d^6\bar{z}\bar{\pi}^{\dot{\alpha}}\bar{j}_{\dot{\alpha}}\nonumber\\
&+&\int d^8zZ J+S[\Phi,\bar{\Phi}],
\eea
where $S[\bar{\Phi},\Phi]$ is the functional responsible for the
dynamics of the chiral matter superfields, while $j_\alpha$ and $J$ are corresponding chiral and real sources, respectively.

In the same way as we described in the last subsection, we can use the Euler-Lagrange equations of $S_M^{(1)}$ either to eliminate the gauge prepotentials from (\ref{master_1}) and obtain 
\bea
\label{sd_1_action}
S_{SD}^{(1)}&=&-\frac{\chi\vartheta}{2}\int d^8z\left(\pi^\alpha D_\alpha Z+\bar{\pi}^{\dot{\alpha}} \bar{D}_{\dot{\alpha}} Z\right)+\frac{1}{4}\left(\int d^6z\pi^\alpha \pi_\alpha+\int d^6\bar z\bar\pi^{\dot{\alpha}}\bar\pi_{\dot{\alpha}}\right)+\frac{m_V^2}{2}\int d^8zV^2\nonumber\\
&+&\int d^6z\pi^\alpha j_\alpha+\int d^6\bar{z}\bar{\pi}^{\dot{\alpha}}\bar{j}_{\dot{\alpha}}+\int d^8zZ J+S[\Phi,\bar{\Phi}].
\eea
or to remove the auxiliary superfields from (\ref{master_1}) and get 
\bea
\label{tm_1_action}
S_{TM}^{(1)}&=&\frac{\vartheta^4}{4}\left(\int d^6zW^\alpha W_\alpha+\int d^6\bar{z}\bar{W}^{\dot{\alpha}}\bar{W}_{\dot{\alpha}}\right)-\frac{\vartheta^2}{2m^2}\int d^8zG^2-\chi\vartheta\int d^8zVG\nonumber\\
&-&\int d^6z\left(i\chi\vartheta W^\alpha+j^\alpha\right)j_\alpha+\int d^6\bar{z}\left(i\chi\vartheta \bar{W}^{\dot{\alpha}}-\bar{j}^{\dot{\alpha}}\right)\bar{j}_{\dot{\alpha}}\nonumber\\
&-&\int d^8z\left(2\chi\vartheta G+J\right)\frac{J}{2m^2}+S[\Phi,\bar{\Phi}].
\eea
Not surprisingly, Eqs. (\ref{sd_1_action}) and (\ref{tm_1_action}) are extensions of the self-dual and topologically massive models which include interactions with matter. Notice that, in contrast to the self-dual model (\ref{sd_1_action}), the topologically massive model (\ref{tm_1_action}) contains a Thirring-like current–current interaction and the gauge prepotentials interact with matter through a non-minimal magnetic-like coupling, as in Refs. \cite{malacarne,Anacleto2001,Ferrari1,Ferrari2}.

On the one hand, the variation of the model (\ref{sd_1_action}) with respect to $Z$, $\pi^\alpha$, and $\bar{\pi}^{\dot{\alpha}}$ leads to the Euler-Lagrange equations
\bea
\label{eq_Z_j}
\frac{\delta S_{SD}^{(1)}}{\delta Z}&=&\frac{\chi\vartheta}{2}\left(D^\alpha\pi_\alpha+\bar{D}^{\dot{\alpha}}\bar{\pi}_{\dot{\alpha}}\right)+m^2Z+J=0  \ ;\\
\label{eq_pi_j}
\frac{\delta S_{SD}^{(1)}}{\delta \pi^\alpha}&=&-\frac{\chi\vartheta}{2}\bar{D}^2D_\alpha Z+\frac{1}{2}\pi_\alpha+j_\alpha=0 \ ;\\
\label{eq_barpi_j}
\frac{\delta S_{SD}^{(1)}}{\delta\bar\pi^{\dot{\alpha}}}&=&-\frac{\chi\vartheta}{2}D^2\bar{D}_{\dot{\alpha}} Z+\frac{1}{2}\bar{\pi}_{\dot{\alpha}}+\bar{j}_{\dot{\alpha}}=0,
\eea
which are just the equations of motion (\ref{eq_Z}-\ref{eq_barpi}) in the presence of external sources. On the other hand, the variation of the action (\ref{tm_1_action}) with respect to $V$, $\psi^\alpha$, and $\bar{\psi}^{\dot{\alpha}}$ leads to the equations of motion
\bea
\label{eq_V_j}
\frac{\delta S_{TM}^{(1)}}{\delta V}&=&-\frac{i\vartheta^2}{2}D^\alpha W_\alpha+\frac{i\vartheta^2}{2}\bar{D}^{\dot{\alpha}} \bar{W}_{\dot{\alpha}}-\chi\vartheta G-\chi\vartheta D^\alpha j_\alpha-\chi\vartheta\bar{D}^{\dot{\alpha}}\bar{j}_{\dot{\alpha}}=0 \ ;\\
\label{eq_psi_j}
\frac{\delta S_{TM}^{(1)}}{\delta \psi^\alpha}&=&-\frac{\vartheta^2}{2m^2}\bar D^2D_\alpha G+i\frac{\chi\vartheta}{2}W_\alpha-\frac{\chi\vartheta}{2m^2}\bar{D}^2D_\alpha J=0 \ ;\\
\label{eq_barpsi_j}
\frac{\delta S_{TM}^{(1)}}{\delta\bar\psi^{\dot{\alpha}}}&=&-\frac{\vartheta^2}{2m^2}D^2\bar D_{\dot{\alpha}} G-i\frac{\chi\vartheta}{2}\bar W_{\dot{\alpha}}-\frac{\chi\vartheta}{2m^2}D^2\bar{D}_{\dot{\alpha}} J=0 \ .
\eea
which are the Euler-Lagrange equations (\ref{eq_V}-\ref{eq_barpsi}) extended to include external source terms.

It is not difficult to see that the equations of motion (\ref{eq_Z_j}-\ref{eq_barpi_j}) and (\ref{eq_V_j}-\ref{eq_barpsi_j}) are equivalent to each other if we make the following generalized identifications:
\bea
Z\longleftrightarrow-\frac{\chi\vartheta}{m^2}G-\frac{1}{m^2}J \ ; \ \pi_\alpha\longleftrightarrow-i\chi\vartheta W_\alpha-2j_\alpha \ ; \ \bar{\pi}_{\dot{\alpha}}\longleftrightarrow i\chi\vartheta\bar{W}_{\dot{\alpha}}-2\bar{j}_{\dot{\alpha}}.
\eea
It is worth to point out that if the currents were simply c-number external sources, we could claim that classical equivalence is maintained even when the superfields are coupled to external sources. However, since the currents are functions of dynamical chiral matter superfields, we need to go a step further, and demonstrate that the matter sectors of the two theories (\ref{sd_1_action}) and (\ref{tm_1_action}) have the same dynamics. To achieve this, let us first consider the variation of the action $S_{SD}^{(1)}$ with respect to the chiral superfields $\Phi$.  It follows from (\ref{sd_1_action}) that
\bea
\label{eq_sd_matter}
\frac{\delta S_{SD}^{(1)}}{\delta \Phi}=0\Rightarrow\frac{\delta S}{\delta \Phi}=-\int d^8z Z\frac{\delta J}{\delta \Phi}-\int d^6z\pi^\alpha\frac{\delta j_\alpha}{\delta \Phi}-\int d^6\bar{z}\bar{\pi}^{\dot{\alpha}}\frac{\delta \bar{j}_{\dot{\alpha}}}{\delta\Phi}.
\eea
We want to solve Eqs. (\ref{eq_Z_j}-\ref{eq_barpi_j}) for $Z$, $\pi^\alpha$, and $\bar{\pi}^{\dot{\alpha}}$ and substitute them into (\ref{eq_sd_matter}). The first step in this direction is to decouple the equations by turning them into second-order differential equations via differentiation and substitution. Doing this for the superfield $Z$ and using the constraints 
\bea
\label{const_j}
D^2 Z = -\frac{D^2 J}{m^2} \ ; \ \bar{D}^2 Z = -\frac{\bar{D}^2 J}{m^2}; \ D^\alpha \pi_\alpha+2 D^\alpha j_\alpha = \bar{D}^{\dot{\alpha}}\bar{\pi}_{\dot{\alpha}}+2 \bar{D}^{\dot{\alpha}}\bar{j}_{\dot{\alpha}},
\eea
which follow directly from (\ref{eq_Z_j}-\ref{eq_barpi_j}), 
we find the following inhomogeneous relativistic wave equation
\bea
\label{wave_Z}
\left(\vartheta^2\Box-m^2\right)Z=\Big(1-\frac{\vartheta^2}{m^2}\{D^2,\bar{D}^2\}\Big)J-\chi\vartheta\left(D^\alpha j_\alpha+\bar{D}^{\dot{\alpha}}\bar{j}_{\dot{\alpha}}\right).
\eea
Since the wave operator $\widehat{\mathcal{O}}^{-1}\equiv\left(\vartheta^2\Box-m^2\right)$ is non-degenerate, there exists an inverse operator $\widehat{\mathcal{O}}$.  Therefore, the solution of (\ref{wave_Z}) can be written as
\bea
\label{solution_Z}
Z=\widehat{\mathcal{O}}\bigg[\Big(1-\frac{\vartheta^2}{m^2}\{D^2,\bar{D}^2\}\Big)J-\chi\vartheta\left(D^\alpha j_\alpha+\bar{D}^{\dot{\alpha}}\bar{j}_{\dot{\alpha}}\right)\bigg].
\eea
The solutions for $\pi^\alpha$ and $\bar{\pi}^{\dot{\alpha}}$ can be found by repeating a similar reasoning that led us to (\ref{solution_Z}). For this reason, we will only present
the final results. The solutions are
\bea
\label{solution_pi}
\pi^\alpha&=&-j^\alpha+\widehat{\mathcal{O}}\left(m^2 j^\alpha-i\vartheta^2\partial^{\alpha\dot{\alpha}}\bar{D}^2\bar{j}_{\dot{\alpha}}+\chi\vartheta\bar{D}^2 D^\alpha J\right);\\
\label{solution_barpi}
\bar{\pi}^{\dot{\alpha}}&=&-\bar{j}^{\dot{\alpha}}+\widehat{\mathcal{O}}\left(m^2\bar{j}^{\dot{\alpha}}-i\vartheta^2\partial^{\alpha\dot{\alpha}}D^2 j_{\alpha}+\chi\vartheta D^2 \bar{D}^{\dot{\alpha}} J\right).
\eea
Plugging (\ref{solution_Z}-\ref{solution_barpi}) into Eq. (\ref{eq_sd_matter}), we obtain
\bea
\label{eq_sd_matter1}
\frac{\delta S}{\delta \Phi}&=&\int d^8z\widehat{\mathcal{O}}\bigg[\Big(\frac{\vartheta^2}{m^2}\{D^2,\bar{D}^2\}-1\Big)J+\chi\vartheta\left(D^\alpha j_\alpha+\bar{D}^{\dot{\alpha}}\bar{j}_{\dot{\alpha}}\right)\bigg]\frac{\delta J}{\delta \Phi}\nonumber\\
&+&\int d^6z\left[j^\alpha-\widehat{\mathcal{O}}\left(m^2 j^\alpha-i\vartheta^2\partial^{\alpha\dot{\alpha}}\bar{D}^2\bar{j}_{\dot{\alpha}}+\chi\vartheta\bar{D}^2 D^\alpha J\right)\right]\frac{\delta j_\alpha}{\delta \Phi}\nonumber\\
&+&\int d^6\bar{z}\left[\bar{j}^{\dot{\alpha}}-\widehat{\mathcal{O}}\left(m^2\bar{j}^{\dot{\alpha}}-i\vartheta^2\partial^{\alpha\dot{\alpha}}D^2 j_{\alpha}+\chi\vartheta D^2 \bar{D}^{\dot{\alpha}} J\right)\right]\frac{\delta \bar{j}_{\dot{\alpha}}}{\delta \Phi}.
\eea
Now, let us find the equations of motion which follow from the variation of the action $S_{TM}^{(1)}$ with respect to the chiral superfields $\Phi$. They are given by
\bea
\label{eq_tm_matter}
\frac{\delta S_{TM}^{(1)}}{\delta \Phi}=0\Rightarrow\frac{\delta S}{\delta \Phi}&=&\int d^8z \frac{1}{m^2}\left(J+\chi\vartheta G\right)\frac{\delta J}{\delta \Phi}+\int d^6z\left(2j^\alpha+i\chi\vartheta W^\alpha\right)\frac{\delta j_\alpha}{\delta \Phi}\nonumber \\
&+&\int d^6\bar{z}\left(2\bar{j}^{\dot{\alpha}}-i\chi\vartheta \bar{W}^{\dot{\alpha}}\right)\frac{\delta \bar{j}_{\dot{\alpha}}}{\delta\Phi}.
\eea
In the same way as we did above, we can decouple the equations (\ref{eq_V_j}-\ref{eq_barpsi_j}) using differentiation and substitution. Proceeding this way, we get inhomogeneous relativistic wave equations for each superfield strength:
\bea
\left(\vartheta^2\Box-m^2\right)G&=&\chi\vartheta D^\alpha\bar{D}^2D_\alpha J+m^2\left(D^\alpha j_\alpha+\bar{D}^{\dot{\alpha}}\bar{j}_{\dot{\alpha}}\right);\\
\left(\vartheta^2\Box-m^2\right)W^\alpha&=&i\chi\vartheta\Box j^\alpha+\chi\vartheta\partial^{\alpha\dot{\alpha}}\bar{D}^2\bar{j}_{\dot{\alpha}}+i\bar{D}^2 D^\alpha J;\\
\left(\vartheta^2\Box-m^2\right)\bar{W}^{\dot{\alpha}}&=&-i\chi\vartheta\Box\bar{j}^{\dot{\alpha}}-\chi\vartheta\partial^{\alpha\dot{\alpha}} D^2 j_{\alpha}-iD^2 \bar{D}^{\dot{\alpha}} J.
\eea
These differential equations have the following solutions
\bea
G&=&-\frac{\chi}{\vartheta}J+\widehat{\mathcal{O}}\left[-\frac{\chi m^2}{\vartheta}J+\chi\vartheta\{D^2,\bar{D}^2\}J+m^2\left(D^\alpha j_\alpha+\bar{D}^{\dot{\alpha}}\bar{j}_{\dot{\alpha}}\right)\right];\\
W^\alpha&=&i\frac{\chi}{\vartheta}j^\alpha+\widehat{\mathcal{O}}\left(i\frac{\chi m^2}{\vartheta}j^\alpha+\chi\vartheta\partial^{\alpha\dot{\alpha}}\bar{D}^2\bar{j}_{\dot{\alpha}}+i\bar{D}^2 D^\alpha J\right);\\
\bar{W}^{\dot{\alpha}}&=&-i\frac{\chi}{\vartheta}\bar{j}^{\dot{\alpha}}-\widehat{\mathcal{O}}\left(i\frac{\chi m^2}{\vartheta} \bar{j}^{\dot{\alpha}}+\chi\vartheta\partial^{\alpha\dot{\alpha}} D^2 j_{\alpha}+iD^2 \bar{D}^{\dot{\alpha}} J\right),
\eea
where we have used the identities
\bea
\Box&=&\frac{1}{\vartheta^{2}}(\widehat{\mathcal{O}}^{-1}+m^2);\\
D^\alpha\bar{D}^2 D_\alpha&=&-\frac{1}{\vartheta^2}(\widehat{\mathcal{O}}^{-1}+m^2)+\left\{D^2,\bar{D}^2\right\}.
\eea
Finally, replacing these solutions in Eq. (\ref{eq_tm_matter}), we obtain
\bea
\label{eq_tm_matter1}
\frac{\delta S}{\delta \Phi}&=&\int d^8z\widehat{\mathcal{O}}\bigg[\Big(\frac{\vartheta^2}{m^2}\{D^2,\bar{D}^2\}-1\Big)J+\chi\vartheta\left(D^\alpha j_\alpha+\bar{D}^{\dot{\alpha}}\bar{j}_{\dot{\alpha}}\right)\bigg]\frac{\delta J}{\delta \Phi}\nonumber\\
&+&\int d^6z\left[j^\alpha-\widehat{\mathcal{O}}\left(m^2 j^\alpha-i\vartheta^2\partial^{\alpha\dot{\alpha}}\bar{D}^2\bar{j}_{\dot{\alpha}}+\chi\vartheta\bar{D}^2 D^\alpha J\right)\right]\frac{\delta j_\alpha}{\delta \Phi}\nonumber\\
&+&\int d^6\bar{z}\left[\bar{j}^{\dot{\alpha}}-\widehat{\mathcal{O}}\left(m^2\bar{j}^{\dot{\alpha}}-i\vartheta^2\partial^{\alpha\dot{\alpha}}D^2 j_{\alpha}+\chi\vartheta D^2 \bar{D}^{\dot{\alpha}} J\right)\right]\frac{\delta \bar{j}_{\dot{\alpha}}}{\delta \Phi}.
\eea
By a direct inspection of Eqs. (\ref{eq_sd_matter}) and (\ref{eq_tm_matter}), we can claim that the matter sectors of the two theories (\ref{sd_1_action}) and (\ref{tm_1_action}) have the same dynamics. This allows us to conclude that the classical equivalence between the supersymmetric self-dual and topologically massive models is maintained even when the superfields are linearly coupled to dynamical matter superfields.

\section{Quantum Equivalence}

It is important to point out that equivalence at the classical level does not necessarily imply equivalence at the quantum level, because in the quantum theory the generating functional is defined by integrating over all possible field configurations, not only the ones which satisfy the equations of motion. Thus, in order to investigate whether the duality holds at the quantum level, we define the generating functional of Green functions in the master theory (\ref{master_1}) as
\bea
\label{z_master}
\mathcal{Z}&=&N\int \mathcal{D}V\mathcal{D}\psi^\alpha\mathcal{D}\bar{\psi}^{\dot{\alpha}}\mathcal{D}Z\mathcal{D}\pi^\alpha\mathcal{D}\bar{\pi}^{\dot{\alpha}}\exp\Bigg\{\frac{1}{4}\left(\int d^6z\pi^\alpha \pi_\alpha+\int d^6\bar z\bar\pi^{\dot{\alpha}}\bar\pi_{\dot{\alpha}}\right)\nonumber\\
&+&i\frac{\chi\vartheta}{2}\bigg(\int d^6z\pi^\alpha W_\alpha-\int d^6\bar z\bar\pi^{\dot{\alpha}}\bar W_{\dot{\alpha}}\bigg)+\int d^8z\left[\frac{m^2}{2} Z^2+\chi\vartheta (Z-V)G\right]+\int d^6z\pi^\alpha j_\alpha\nonumber\\
&+&\int d^6\bar{z}\bar{\pi}^{\dot{\alpha}}\bar{j}_{\dot{\alpha}}+\int d^8zZ J+S[\Phi,\bar{\Phi}]\Bigg\},
\eea
where $N$ is a normalization constant which will be used to absorb field-independent factors. 

On the one hand, it is convenient to change the functional integration variables in (\ref{z_master}) to $V\rightarrow V+Z$, $\psi^\alpha\rightarrow\psi^\alpha-\pi^\alpha$, and $\bar{\psi}^{\dot{\alpha}}\rightarrow\bar{\psi}^{\dot{\alpha}}-\bar{\pi}^{\dot{\alpha}}$. Since we assume that the superfields $Z$, $\pi^\alpha$, and $\bar{\pi}^{\dot{\alpha}}$ do not change under these transformations, the Jacobian of such a change of variables is equal to one. Proceeding this way, the superfields $V$, $\psi^\alpha$, and $\bar{\psi}^{\dot{\alpha}}$ completely decouple and the functional integration over these superfields yields the following result
\bea
\label{z_SD1}
\mathcal{Z}=N\int\mathcal{D}Z\mathcal{D}\pi^\alpha\mathcal{D}\bar{\pi}^{\dot{\alpha}}\exp\left(S_{SD}^{(1)}\right),
\eea
where $S_{SD}^{(1)}$ is the model obtained in (\ref{sd_1_action}). 

On the other hand, let us consider this time the following transformations in (\ref{z_master}): $Z\rightarrow Z-\frac{\chi\vartheta}{m^2}G-\frac{1}{m^2}J$, $\pi^\alpha\rightarrow\pi^\alpha-i\chi\vartheta W^\alpha-2j^\alpha$, and $\bar{\pi}^{\dot{\alpha}}\rightarrow\bar{\pi}^{\dot{\alpha}}+i\chi\vartheta \bar{W}^{\dot{\alpha}}-2\bar{j}^{\dot{\alpha}}$. Since these transformations are simply shifts by a constant, they leave the integration measures in the functional integrals invariant and decouple the superfields $Z$, $\pi^\alpha$, and $\bar{\pi}^{\dot{\alpha}}$. This allows us to easily perform integrations over $Z$, $\pi^\alpha$, and $\bar{\pi}^{\dot{\alpha}}$ and obtain
\bea
\label{z_TM1}
\mathcal{Z}=N\int\mathcal{D}V\mathcal{D}\psi^\alpha\mathcal{D}\bar{\psi}^{\dot{\alpha}}\exp\left(S_{TM}^{(1)}\right),
\eea
where $S_{TM}^{(1)}$ is the action found in (\ref{tm_1_action}).

Notice that (\ref{z_SD1}) and (\ref{z_TM1}) are generating functionals in the supersymmetric self-dual and topologically massive models, respectively. The fact that there exists a master generating functional (\ref{z_master}) which interpolates between (\ref{z_SD1}) and (\ref{z_TM1}) is a strong evidence that the duality holds at the quantum level. However, in order to complete the proof of the quantum equivalence between the supersymmetric self-dual and topologically massive models, we need to carry out the remaining integrals in (\ref{z_SD1}) and (\ref{z_TM1}).

First, let us consider the generating functional (\ref{z_SD1}). By using Eq. (\ref{sd_1_action}), we can rewrite (\ref{z_SD1}) in a more convenient form:
\bea
\label{z_SD2}
\mathcal{Z}&=&N e^{S[\bar{\Phi},\Phi]}\int\mathcal{D}Z\exp\int d^8z\left(\frac{m}{2}Z^2+ZJ\right)\int\mathcal{D}\pi^\alpha\exp\int d^6z\bigg[\frac{1}{4}\pi^\alpha\pi_\alpha+\pi^\alpha\bigg(-\frac{\chi\vartheta}{2}\bar{D}^2 D_\alpha Z\nonumber\\
&+&j_\alpha\bigg)\bigg]\int\mathcal{D}\bar{\pi}^{\dot{\alpha}}\exp\int d^6\bar{z}\left[\frac{1}{4}\bar{\pi}^{\dot{\alpha}}\bar{\pi}_{\dot{\alpha}}+\bar{\pi}^{\dot{\alpha}}\left(-\frac{\chi\vartheta}{2}D^2 \bar{D}_{\dot{\alpha}} Z+\bar{j}_{\dot{\alpha}}\right)\right].
\eea
We see that the integrals over the superfields $\pi^\alpha$ and $\bar{\pi}^{\dot{\alpha}}$ are ordinary Gaussian integrals that can be evaluated directly. Therefore, we find
\bea
\label{z_SD3}
\mathcal{Z}&=&N \exp\left(S[\bar{\Phi},\Phi]-\int d^6zj^\alpha j_\alpha-\int d^6\bar{z}\bar{j}^{\dot{\alpha}} \bar{j}_{\dot{\alpha}}\right)\int\mathcal{D}Z\exp\int d^8z\bigg[\frac{1}{2}Z\big(\vartheta^2D^\alpha\bar{D}^2D_\alpha\nonumber\\
&+&m^2\big)Z+Z\left(J-\chi\vartheta D^\alpha j_\alpha-\chi\vartheta\bar{D}^{\dot{\alpha}}\bar{j}_{\dot{\alpha}}\right)\bigg].
\eea
In order to evaluate the last integral, we first have to find the inverse of the differential operator in the quadratic part in $Z$. It is not hard to show that
\bea
\label{inverse_Z}
\left(\vartheta^2 D^\alpha\bar{D}^2D_\alpha+m^2\right)^{-1}=-\widehat{\mathcal{O}}\left(1-\frac{\vartheta^2}{m^2}\{D^2,\bar{D}^2\}\right).
\eea
Thus, integrating over the superfield $Z$, we are lead to the following result 
\bea
\label{final_SD}
\mathcal{Z}&=&N\exp\bigg\{S[\bar{\Phi},\Phi]+\int d^8z\bigg[\frac{1}{2}J\widehat{\mathcal{O}}\left(1-\frac{\vartheta^2}{m^2}\{D^2,\bar{D}^2\}\right)J-\chi\vartheta J\widehat{\mathcal{O}}\left(D^\alpha j_\alpha+\bar{D}^{\dot{\alpha}}\bar{j}_{\dot{\alpha}}\right)\nonumber\\
&+&i\vartheta^2 j_\alpha\widehat{\mathcal{O}}\partial^{\alpha\dot{\alpha}}\bar{j}_{\dot{\alpha}}\bigg]+\frac{1}{2}\int d^6zj^\alpha\left(m^2\widehat{\mathcal{O}}-1\right)j_\alpha+\frac{1}{2}\int d^6\bar{z}\bar{j}^{\dot{\alpha}}\left(m^2\widehat{\mathcal{O}}-1\right)\bar{j}_{\dot{\alpha}}\bigg\}.
\eea
Now, let us consider the functional integrals in (\ref{z_TM1}). These integrals are ill-defined due to the gauge invariance of the classical action $S_{TM}$. Thus, to make progress, it is necessary to fix the gauge by adding to $S_{TM}^{(1)}$ some gauge-fixing functional $S_{GF}$. We choose \cite{GGRS}
\bea
\label{gaugefixing}
S_{GF}=-\frac{\vartheta^2}{\alpha}\int d^8z\left(\bar{D}^2V\right)D^2V-\frac{\vartheta^2}{8\xi m^2}\int d^8z\left(D^\alpha\psi_\alpha-\bar{D}^{\dot{\alpha}}\bar{\psi}_{\dot{\alpha}}\right)^2,
\eea
where $\alpha$ and $\xi$ are gauge-fixing parameters. Since the Faddeev-Popov ghosts completely decouple for this gauge-fixing functional, their action will be omitted.

Therefore, it follows from (\ref{tm_1_action}), (\ref{z_TM1}), and (\ref{gaugefixing}) that
\bea
\label{z_TM2}
\mathcal{Z}&=&N\int\mathcal{D}V\mathcal{D}\psi^\alpha\mathcal{D}\bar{\psi}^{\dot{\alpha}}\exp\left(S_{TM}^{(1)}+S_{GF}\right)\nonumber\\
&=&N\exp\left(S[\bar{\Phi},\Phi]-\int d^6zj^\alpha j_\alpha-\int d^6\bar{z}\bar{j}^{\dot{\alpha}}\bar{j}_{\dot{\alpha}}-\frac{1}{2m^2}\int d^8zJ^2\right)\nonumber\\
&\times&\int\mathcal{D}V\exp\int d^8z\left[-\frac{\vartheta^2}{2}V\Box\left(\Pi_{\frac{1}{2}}+\frac{1}{\alpha}\Pi_0\right)V-\chi\vartheta V\left(D^\alpha j_\alpha+\bar{D}^{\dot{\alpha}}\bar{j}_{\dot{\alpha}}\right)\right]\nonumber\\
&\times&\int\mathcal{D}\psi^\alpha\exp\int d^6z\left[-\frac{\vartheta^2}{4m^2}\psi^\alpha\Box\left(\Pi_{+}+\frac{1}{\xi}\Pi_-\right)\psi_\alpha-i\frac{\chi\vartheta}{2m^2}\psi^\alpha\bar{D}^2D_\alpha\left(m^2V+J\right)\right]\nonumber\\
&\times&\int\mathcal{D}\bar{\psi}^{\dot{\alpha}}\exp\int d^6\bar{z}\left[-\frac{\vartheta^2}{4m^2}\bar{\psi}^{\dot{\alpha}}\Box\left(\Pi_{+}+\frac{1}{\xi}\Pi_-\right)\bar{\psi}_{\dot{\alpha}}-i\frac{\chi\vartheta}{2m^2}\bar{\psi}^{\dot{\alpha}}D^2\bar{D}_{\dot{\alpha}}\left(m^2V+J\right)\right], 
\eea
where we have introduced two sets of projection operators. The first set of operators are the familiar projectors on the transverse and longitudinal parts of the superfield $V$:
\bea
\Pi_{\frac{1}{2}}=-\frac{D^\alpha\bar{D}^2D_\alpha}{\Box} \ ; \ \Pi_0=\frac{\{D^2,\bar{D}^2\}}{\Box}.
\eea
The second set of operators are defined in terms of the rest-frame conjugation operator ${\bf K}$ \cite{SG,GGRS}:
\bea
\Pi_{\pm}=\frac{1}{2}(1\pm {\bf K}),
\eea
where ${\bf K}$ acts on a $\mathcal{N}=1$ chiral spinor superfield $\psi_\alpha$ in the following way
\bea
{\bf K}\psi_\alpha=-\frac{\bar{D}^2i{\partial_\alpha}^{\dot{\alpha}} }{\Box}\bar{\psi}_{\dot{\alpha}}.
\eea
Of course, both sets of projection operators satisfy relations of completeness, idempotence, and orthogonality. Due to these properties, projection operators make the task of inverting the differential operators in (\ref{z_TM2}) quite easy. For example, it is trivial to show that
\bea
\left[-\frac{\vartheta^2}{2m^2}\Box\left(\Pi_++\frac{1}{\xi}\Pi_-\right)\right]^{-1}=-\frac{2m^2}{\vartheta^2\Box}\left(\Pi_++\xi\Pi_-\right).
\eea
This propagator allows us to integrate (\ref{z_TM2}) over $\psi^\alpha$ and $\bar{\psi}^{\dot{\alpha}}$. After these integrations, we can manipulate the results, with the help of the identities
\bea
\label{Pi_+}
\Pi_+\bar{D}^2D_\alpha\left(m^2V+J\right)=\bar{D}^2D_\alpha\left(m^2V+J\right)&;&\Pi_+D^2\bar{D}_{\dot{\alpha}}\left(m^2V+J\right)=D^2\bar{D}_{\dot{\alpha}}\left(m^2V+J\right);\\
\label{Pi_-}
\Pi_-\bar{D}^2D_\alpha\left(m^2V+J\right)=0&;&\Pi_-D^2\bar{D}_{\dot{\alpha}}\left(m^2V+J\right)=0,
\eea
and show that
\bea
\label{z_TM3}
\mathcal{Z}&=&N\exp\left[S[\bar{\Phi},\Phi]-\int d^6zj^\alpha j_\alpha-\int d^6\bar{z}\bar{j}^{\dot{\alpha}}\bar{j}_{\dot{\alpha}}+\frac{1}{2m^2}\int d^8zJ\left(\Pi_{\frac{1}{2}}-1\right)J\right]\\
&\times&\int\mathcal{D}V\exp\int d^8z\left[-\frac{1}{2}V\left(\widehat{\mathcal{O}}^{-1}\Pi_{\frac{1}{2}}+\frac{\vartheta^2\Box}{\alpha}\Pi_0\right)V+V\left(\Pi_{\frac{1}{2}}J-\chi\vartheta D^\alpha j_\alpha-\chi\vartheta\bar{D}^{\dot{\alpha}}\bar{j}_{\dot{\alpha}}\right)\right]. \nonumber
\eea
It is also trivial to get the $V$-propagator. It is given by
\bea
\left(-\widehat{\mathcal{O}}^{-1}\Pi_{\frac{1}{2}}-\frac{\vartheta^2\Box}{\alpha}\Pi_0\right)^{-1}=-\left(\widehat{\mathcal{O}}\Pi_{\frac{1}{2}}+\frac{\alpha}{\vartheta^2\Box}\Pi_0\right).
\eea
By means of this propagator we can evaluate the last integration. Then, by making use of
\bea
\label{Pi_1/2}
\Pi_\frac{1}{2}\left(\Pi_{\frac{1}{2}}J-\chi\vartheta D^\alpha j_\alpha-\chi\vartheta\bar{D}^{\dot{\alpha}}\bar{j}_{\dot{\alpha}}\right)&=&\Pi_{\frac{1}{2}}J-\chi\vartheta D^\alpha j_\alpha-\chi\vartheta\bar{D}^{\dot{\alpha}}\bar{j}_{\dot{\alpha}};\\
\label{Pi_0}
\Pi_0\left(\Pi_{\frac{1}{2}}J-\chi\vartheta D^\alpha j_\alpha-\chi\vartheta\bar{D}^{\dot{\alpha}}\bar{j}_{\dot{\alpha}}\right)&=&0,
\eea
we can finally obtain
\bea
\label{final_TM}
\mathcal{Z}&=&N\exp\bigg\{S[\bar{\Phi},\Phi]+\int d^8z\bigg[\frac{1}{2}J\widehat{\mathcal{O}}\left(1-\frac{\vartheta^2}{m^2}\{D^2,\bar{D}^2\}\right)J-\chi\vartheta J\widehat{\mathcal{O}}\left(D^\alpha j_\alpha+\bar{D}^{\dot{\alpha}}\bar{j}_{\dot{\alpha}}\right)\nonumber\\
&+&i\vartheta^2 j_\alpha\widehat{\mathcal{O}}\partial^{\alpha\dot{\alpha}}\bar{j}_{\dot{\alpha}}\bigg]+\frac{1}{2}\int d^6zj^\alpha\left(m^2\widehat{\mathcal{O}}-1\right)j_\alpha+\frac{1}{2}\int d^6\bar{z}\bar{j}^{\dot{\alpha}}\left(m^2\widehat{\mathcal{O}}-1\right)\bar{j}_{\dot{\alpha}}\bigg\}.
\eea
Notice that the $\xi$- and $\alpha$-dependent contributions to the generating functional $\mathcal{Z}$ vanished due to the identities (\ref{Pi_-}) and (\ref{Pi_0}), which have led to a gauge-independent final result (\ref{final_TM}).

We see that the integrations over $Z$, $\pi^\alpha$, and $\bar{\pi}^{\dot{\alpha}}$ in (\ref{z_SD1}) and the integrations over $V$, $\psi^\alpha$, and $\bar{\psi}^{\dot{\alpha}}$ in Eq. (\ref{z_TM1}) resulted in the same effective nonlocal action [see Eqs. (\ref{final_SD}) and (\ref{final_TM})]
\bea
S_{eff}&=&S[\bar{\Phi},\Phi]+\int d^8z\bigg[\frac{1}{2}J\widehat{\mathcal{O}}\left(1-\frac{\vartheta^2}{m^2}\{D^2,\bar{D}^2\}\right)J-\chi\vartheta J\widehat{\mathcal{O}}\left(D^\alpha j_\alpha+\bar{D}^{\dot{\alpha}}\bar{j}_{\dot{\alpha}}\right)\nonumber\\
&+&i\vartheta^2 j_\alpha\widehat{\mathcal{O}}\partial^{\alpha\dot{\alpha}}\bar{j}_{\dot{\alpha}}\bigg]+\frac{1}{2}\int d^6zj^\alpha\left(m^2\widehat{\mathcal{O}}-1\right)j_\alpha+\frac{1}{2}\int d^6\bar{z}\bar{j}^{\dot{\alpha}}\left(m^2\widehat{\mathcal{O}}-1\right)\bar{j}_{\dot{\alpha}}.
\eea
It is not hard to check that the variation of $S_{eff}$ with respect to the matter superfields leads to equations of motion which are exactly those ones found in the previous section, namely Eqs. (\ref{eq_sd_matter}) and (\ref{eq_tm_matter}). Accordingly, we can conclude that there exists a quantum equivalence between supersymmetric self-dual and topologically massive models linearly coupled to dynamical matter.

\section{Summary}

We proved duality between two four-dimensional superfield theories describing the spinor chiral superfield, that is, the supersymmetric self-dual and topologically massive models, both for free and coupled cases, at classical and quantum levels. The importance of our result is motivated by the fact that while, up to now, the duality between self-dual and gauge theories originally proposed in \cite{MCS auto dual} was treated as a typically three-dimensional phenomenon except of a few papers, that is, \cite{BF_dual,Clovis}, we not only promoted duality to the four-dimensional space-time but demonstrated its possibility for supersymmetric field theories defined in this space-time. It worth mentoning also that while most of papers devoted to duality between various versions of self-dual and massive gauge theories considered only the tree-level aspects of the duality, we studied also its quantum manifestation, generalizing results of \cite{Ferrari1,Ferrari2}  to four-dimensional theories. Moreover, we obtained new results for the chiral spinor superfield, which is the less studied one among all superfields listed in \cite{GGRS}, so that only a few its perturbative studies were performed up to now \cite{Almeida,firstBF}.

Certainly, the natural question consists in possible generalizations and extensions of our results. In this context, one way can consist in the analysis of other couplings of the spinor chiral superfield (it is worth mentioning that there are different couplings of the antisymmetric tensor field, see f.e. \cite{Mariz2019}, and their promotion to the superfield level is a very interesting problem). Another line can consist in study of various applications of the duality we demonstrated, within different contexts, from condensed matter to string theory. Besides, certainly it is interesting to study whether it is possible to establish duality between supersymmetric theories describing other supermultiplets discussed in \cite{GGRS}, and extend the duality we proved to the case of a curved superspace generalizing the results obtained in \cite{BuchKuz}. We plan to consider these problems in forthcoming papers.

{\bf Acknowledgments.} The work by A. Yu. P. has been partially supported by the CNPq project No. 301562/2019-9. P. J. P. would like to thank the Brazilian agency CAPES for financial support (PNPD/CAPES grant, process 88887.464556/2019-00).

\end{document}